\documentstyle[sprocl,epsfig]{article}

\def\gsim{\,\lower.25ex\hbox{$\scriptstyle\sim$}\kern-1.30ex%
\raise 0.55ex\hbox{$\scriptstyle >$}\,}
\def\lsim{\,\lower.25ex\hbox{$\scriptstyle\sim$}\kern-1.30ex%
\raise 0.55ex\hbox{$\scriptstyle <$}\,}

\newcommand{\pom}{I\!\!P}
\newcommand{\xpom}{x_{\pom}}

\newcommand{\fiidiiii}{$F_2^{D(4)}$}

\newcommand{\rps}{$\rm Roman~pots$ }

\begin{document}

\title{PROPOSAL FOR A VERY FORWARD PROTON SPECTROMETER IN H1 AFTER
2000}

\author{L. FAVART} 

\address{Universit\'e Libre de Bruxelles, IIHE \\
        on behalf of the H1 Collaboration
        \\E-mail: lfavart@ulb.ac.be} 


\maketitle\abstracts{ 
A new, very forward proton spectrometer (VFPS) with large acceptance is
proposed to
be installed in the proton beam of the H1 experiment after the
luminosity upgrade in the year 2000. The spectrometer, located
at 220~m downstream of the interaction point is based on the Roman Pot
technique and consists of two stations situated in the cold section
of the proton beam line. Physics motivations and a brief
description of the proton spectrometer are presented.
}

\section{Physics goal}

In recent years, 
due to the results obtained by the ZEUS and H1 experiments
at HERA, considerable
progress has been achieved in the partonic interpretations of 
diffractive processes, see e.g.~\cite{arneodo}. 
However, the small cross sections involved and the difficulty in selecting 
clean diffractive event samples have left many basic QCD predictions 
untested. Therefore further progress in this field will rely on collecting
large statistics in various inclusive, semi-inclusive and exclusive 
diffractive channels, in particular those in which a hard scale is
present.
\\

Most of diffractive studies performed up to now at HERA have been based on 
the characteristic presence of a rapidity gap in the diffractive final state.
However the only precise and unambiguous
way of studying diffraction is by
tagging the diffracted proton and measuring its four momentum
by means of a proton spectrometer.
Such devices have been used by the H1 and ZEUS Collaborations
and have delivered interesting results, 
but their acceptances are small, with the result that the collected 
statistics are limited and large systematic errors affect the
measurements.
To fully profit from the HERA luminosity upgrade in the 
study of diffraction after the year 2000, 
a proton spectrometer which identifies and 
measures the momentum of the diffracted proton with a very
good acceptance is thus essential.
\\

\begin{figure}[htbp]
 \vskip 1.0cm
 \begin{picture}(100,100)
 \put(25,0){\epsfig{figure=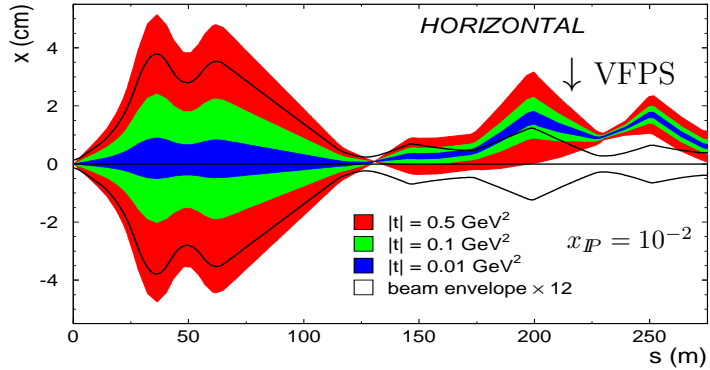,height=5.0cm,width=0.85\textwidth}}
  \put(235,110){ \Large $\downarrow$ \large VFPS}
  \put(240,48){$\xpom = 10^{-2}$}
 \end{picture}
 \caption{Horizontal projections of the beam envelope (12 sigma), 
  as a function of the distance to the interaction point
  (hatched areas).
  The projection of the transverse distances of the scattered protons for
  three different $t$ values and for $\xpom=10^{-2}$ are given by the 
  shaded areas.
  The vertical arrow indicates the position for the new \rps. }
 \label{fig:beam}
\end{figure}

\begin{figure}[htbp]
 \begin{picture}(100,100)
  \put(-5,0){\epsfig{figure=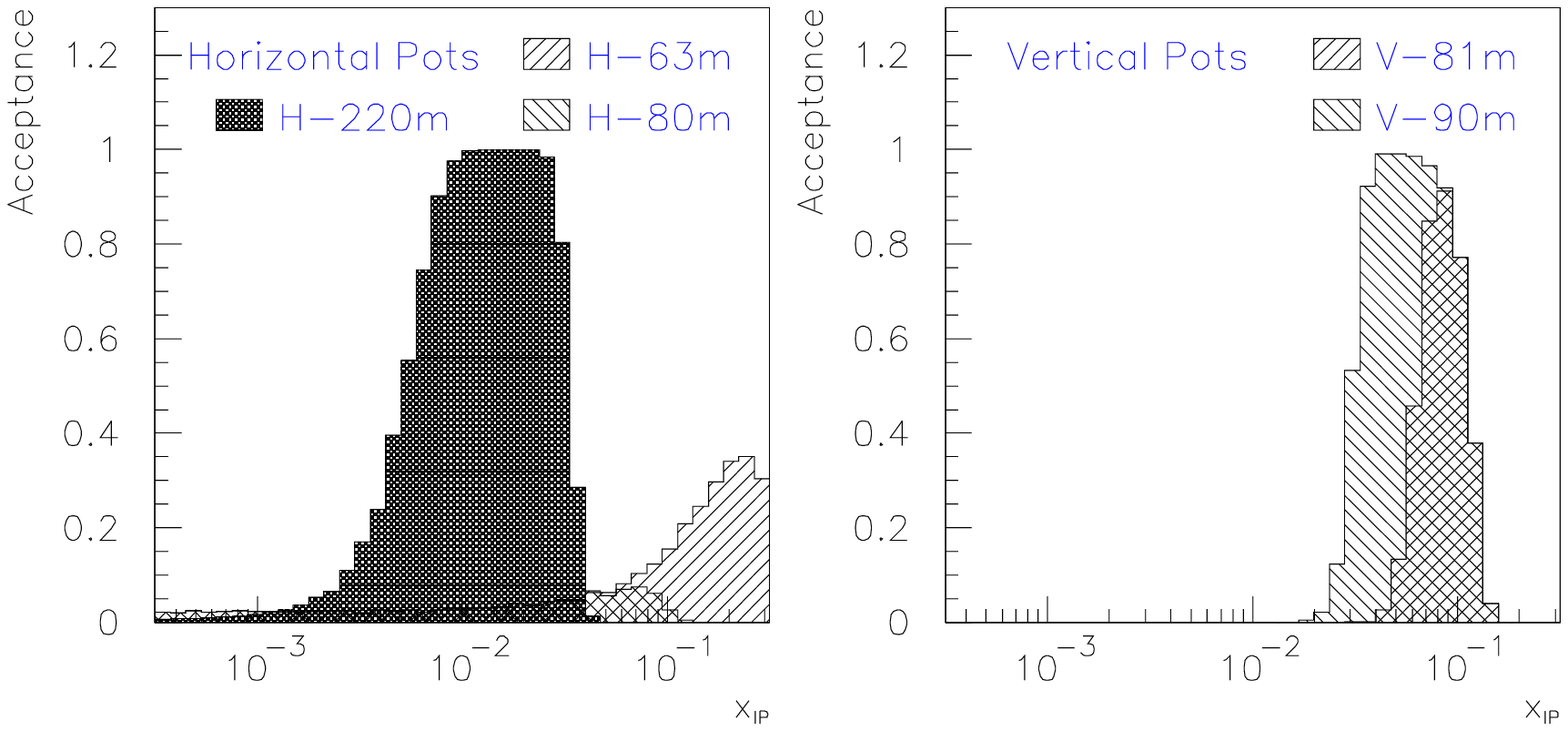,height=4.cm}}
  \put(235,12){\epsfig{figure=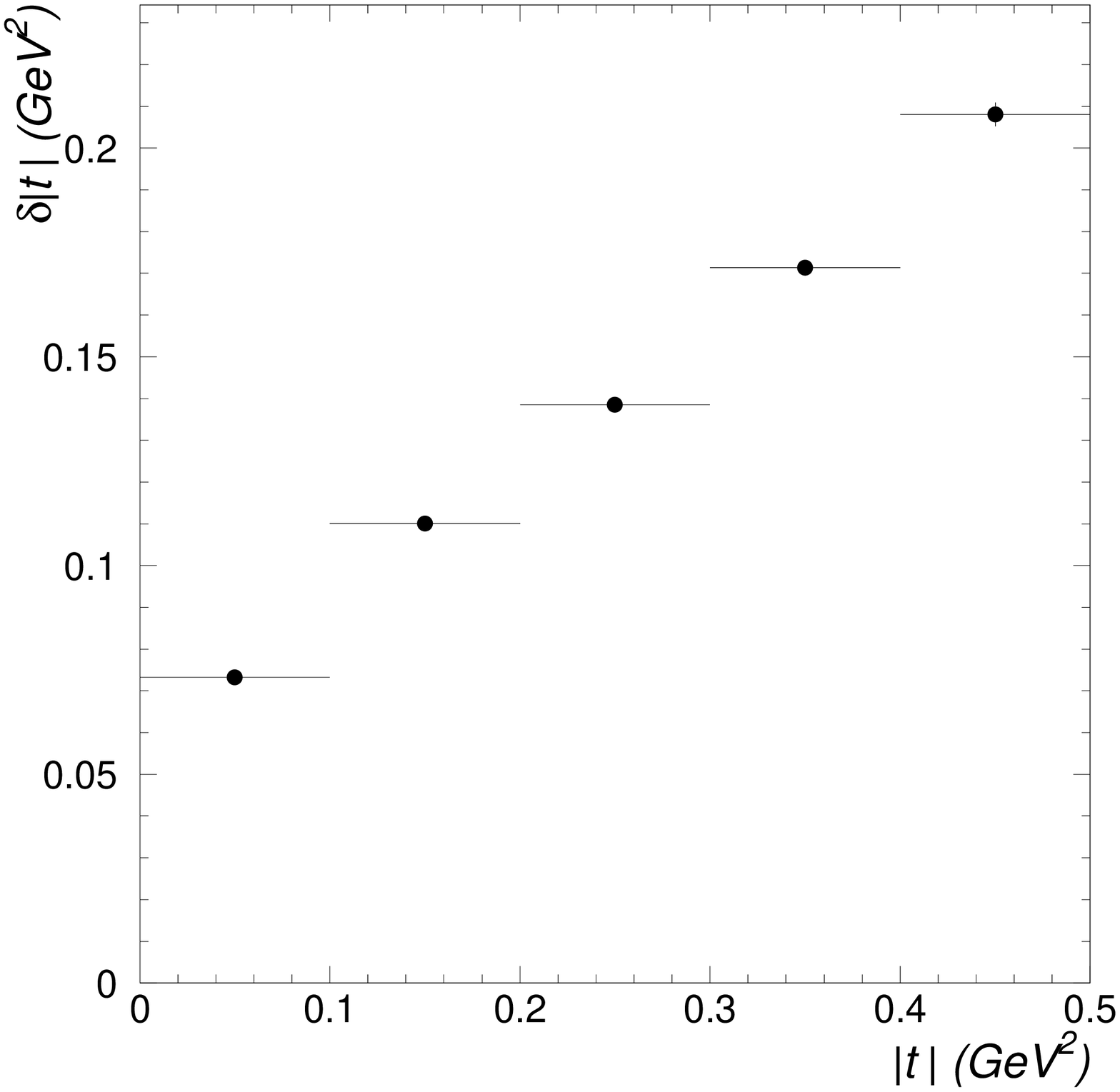,height=3.6cm,width=3.8cm}}
  \put(95,85){\bf a}
  \put(210,84){\bf b}
  \put(330,85){\bf c}
 \end{picture}
 \caption{a) and b): acceptances of all present and proposed (220m) FPS as a 
  function of $\xpom$ for the vertical and horizontal stations respectively.
  c) resolution of $t$ as a function of $t$.}
 \label{fig:acc}
\end{figure}

The installation of a new proton spectrometer is 
proposed\footnote{Antwerp (UIA), Birmingham, Brussels (ULB-VUB), 
Hamburg II and Lund groups from the H1 Collaboration}
at 220 m downstream of the H1 main detector.
In the proposed location, the strong horizontal beam bend,
as shown in Fig.~\ref{fig:beam}, allows scattered protons to be measured
down to the lowest $|t|$ values\footnote{$t$ beeing the four-momentum
transfer squared at the proton vertex}: 
$|t_{min}| < |t| \, \lsim \, 0.5$ GeV$^2$.
The anticipated acceptance, shown in Fig.~\ref{fig:acc}a) and b) as a function
of $\xpom$, is above 80 \% for $5.10^{-3} < \xpom < 3.10^{-2}$.
With the measurement of two impact point positions, 
the variables $\xpom$, $t$ and the azimuthal angle of the 
scattered proton can be determined. 
The resolution in $t$, shown in Fig.~\ref{fig:acc}c),
permits a measurement of 4-5 bins in $t$~\cite{prc}. 
\\

The installation of the high acceptance proton 
spectrometer will thus provide remarkable improvements 
in the study of diffraction: 
genuine elastic events not contaminated by proton dissociation 
can be selected with high acceptance, in particular, 
for high transverse momentum jets and charm analysis.
On top of that, 
measurements of basic importance can be performed: determination of the 
longitudinal cross section using the measurement of the azimuthal angle of 
the scattered proton and measurement of the fully differential
\fiidiiii structure function, including its $t$-dependence down to the
lowest $t$ values.

\section{Roman Pot detectors}

 The proton spectrometer (PS) is a set of two ``Roman pots''.
Each pot consists of an insert into the beam pipe,
allowing two tracking detectors equipped with scintillating fibres
to be moved very close to the proton beam.
\\

The Roman pots will be installed in the
``cold'' part of the HERA proton ring,
which is equipped with superconducting magnets.
However, in order to access the beam pipe with Roman pot detectors, the proton
beam line has to be at room temperature. 
This implies that the beam pipe has to be separated in this area 
from the cold elements of the drift tube, 
and a bypass has to be installed to transport horizontally the
helium lines to the next cold section~\cite{prc}.
\\

Many aspects of the design of the Roman pots, including the
stainless plunger vessel and the scintillating fibre detectors,
are adaptations of the existing proton spectrometer, FPS~\cite{fps}, 
installed and operational in H1 since 1994. 
Both detectors of each Roman pot consists of two planes of 
scintillating fibres perpendicular to the beam line direction and
oriented at $\pm 45^0$ from the horizontal direction.
Each detector thus provides the reconstruction of the position of one 
impact point 
of the scattered proton trajectory with a precision of about 100 $\mu$m,
leading (after inclusion of the beam spread effect) to the resolution shown
above.

 The installation should take place during 2001-2002, so that data taking
with the new proton spectrometer will start in 2003.

\end{document}